# Spatiotemporal beam self-cleaning for high-resolution nonlinear fluorescence imaging with multimode fibres

**Nawell Ould Moussa[1], Tigran Mansuryan[1], Charles Henri Hage[1], Marc Fabert[1], Katarzyna Krupa[2], Alessandro Tonello[1], Mario Ferraro[3], Luca Leggio[3], Mario Zitelli[3], Fabio Mangini[4], Alioune Niang[4], Guy Millot[5,6], Massimiliano Papi[7], Stefan Wabnitz[3,8], and Vincent Couderc[1,*]**

[1] *Université de Limoges, XLIM, UMR CNRS 7252, 123 Avenue A. Thomas, 87060 Limoges, France*
[2] *Institute of Physical Chemistry, Polish Academy of Sciences, ul. Kasprzaka 44/52, 01-224 Warsaw, Poland*
[3] *DIET, Sapienza University of Rome Via Eudossiana 18, 00184 Rome, Italy*
[4] *Dipartimento di Ingegneria dell'Informazione, Università di Brescia, via Branze 38, 25123, Brescia, Italy*
[5] *Université de Bourgogne Franche-Comté, ICB, UMR CNRS 6303, 9 Avenue A. Savary, 21078 Dijon, France*
[6] *Institut Universitaire de France (IUF), 1 rue Descartes, 75005 Paris, France*
[7] *Dipartimento di Neuroscienze, Università Cattolica del Sacro Cuore, 00168 Rome, Italy*
[8] *Physics Department, Novosibirsk State University, Pirogova 1, Novosibirsk 630090, Russia*

e-mail* Vincent.couderc@xlim.fr

**Beam self-cleaning (BSC) in graded-index (GRIN) multimode fibres (MMFs) has been recently reported by different research groups. Driven by the interplay between Kerr effect and beam self-imaging, BSC counteracts random mode coupling, and forces laser beams to recover a quasi-single mode profile at the output of GRIN fibres. Here we show that the associated self-induced spatiotemporal reshaping allows for improving the performances of nonlinear fluorescence microscopy and endoscopy using multimode optical fibres. We experimentally demonstrate that the beam brightness increase, induced by self-cleaning, enables two and three-photon imaging of biological samples with high spatial resolution. Temporal pulse shortening accompanying spatial beam clean-up enhances the output peak power, hence the efficiency of nonlinear imaging. We also show that spatiotemporal supercontinuum generation is well-suited for large-band nonlinear fluorescence imaging in visible and infrared domains. We substantiated our findings by multiphoton fluorescence imaging in both microscopy and endoscopy configurations.**

Besides their relevance as a test-bed for fundamental research, MMFs have recently attracted a strong attention for their technological applications. For instance, MMFs may provide a solution to the data capacity bottleneck of optical communication links via spatial-division-multiplexing [1-2], and scale-up the output energy of fibre lasers sources [3-5]. Linear and nonlinear imaging in microscopy and endoscopy configurations [6-7] also provide an interesting application of MMFs. Several systems have already been proposed, which all require a careful control of multimode beam propagation. The first images were obtained in a wide-field configuration, based on holographic techniques [8-10]. More recently, holograms have been replaced by spatial light modulators. These permit to shape and control light at the fibre input, and counteract the deleterious distortions undergone by the beam, when carried by the multiple fibre modes. These methods rely on using the inverse matrix of the recorded transmission matrix of the fibre, in order to project the desired output pattern at the fibre end [11-14]. The main serious drawback of these wide-field imaging systems is that any slight fibre movement changes the random mode coupling process. In turn, this strongly modifies the fibre transmission matrix, which thus needs to be updated by performing additional measurements. Moreover, such initial learning stage is time-consuming, which leads to significant slowing down of the imaging rate. A deep neural network approach, capable of learning the input-output relationship of MMFs, has been used to reconstruct images at the fibre output [15-17]. The first learning stage can also be used to transmit other images (not used for training), which may increase the overall robustness of the imaging system. So far, both linear and nonlinear distortions caused by propagation in MMFs have been taken into account, which allowed for the use of short pulses in the formation of nonlinear fluorescence images. Although two-photon microscopy is an imaging technique largely exploited over the last thirty years [18], it has been recently revisited, thanks to the introduction of spatial wavefront shaping. In this way it is possible to achieve high-resolution focusing and tight spatial sectioning even when the light has passed through a dispersive scattering medium. The spatial wavefront shaping permits the pre-compensation of the disorder experienced along the propagation, so that one can optimize both the focusing point [19-20] and the pulse duration [21-22] of the output beam. In this way one can combine both high spatial resolution and efficient nonlinear imaging; an additional scanning of the sample is used to obtain the images. However, the use of spatial wavefront shaping for beam optimization at the MMF output suffers from the same drawbacks than the aforementioned wide-field imaging technique. Namely, the approach requires a non-negligible time to learn, and has a strong sensitivity to random mode coupling variations, thus weakening the performances of the imaging system.

The phenomenon of spatial BSC, which appears spontaneously when energetic light pulses propagate in GRIN MMFs [23-29], was recently employed for 3D high-resolution photoacoustic endoscopy [7]. In that experiment, only the spatial reshaping of the beam was used for increasing the spatial resolution of a photoacoustic endoscope. However, in order to push the performance of nonlinear imaging systems to their limits, the full spatiotemporal dynamics of the self-cleaning process can be exploited. This is the purpose of this Letter: we demonstrate that the spatiotemporal character of BSC leads to significant resolution enhancement in multispectral multiphoton fluorescence imaging, in both microscopy and endoscopy configurations. This is obtained thanks to BSC-induced energetic broadband frequency conversion, activated by significant temporal pulse narrowing, and high-energy multimode soliton generation [24-25, 30-32, 33].

In our experiments (see figure 1a), we coupled a Gaussian pulsed (80 ps) beam at 1064 nm into 3-m or 18-m spans of GRIN MMF, followed by a multiphoton microscope (Thorlabs Bergamo). The microscope was composed of a scanning system, a microscope objective (LUMFLN60XW, Olympus), a set of band-pass filters coupled with dichroic mirrors, and four photomultiplier tubes (see Methods) placed before and after the MMF, in order to measure the emitted fluorescence in microscopy or endoscopy configurations.

By measuring its spatiotemporal features, we first characterized the BSC effect that occurs in the GRIN MMF. In Fig.1b we show the evolution of the output spatial beam diameter with respect to the input peak power. As we can see, spatial self-cleaning occurs at peak pulse powers above 10 kW, leading to a typical bell-shaped beam pattern, surrounded by a low-energy speckled background (see supplementary material). In the temporal domain, we measured a three-fold pulse narrowing; the autocorrelation traces and the corresponding pulse durations assuming a $sech^2$ are shown in figure 1c. Thus, BSC simultaneously improves the output beam quality, and doubles the output pulse peak power. By further increasing the peak pulse power, the initially nearly monochromatic spatiotemporal beam shaping turns into a multicolour one, because of the self-phase-modulation induced spectral broadening. Fig.1d shows that a supercontinuum extending from 0.6 μm up to 2.4 μm is generated at the fibre output. As the pulse power is increased above the threshold for BSC, we observe the so-called geometric parametric instability (GPI) effect, leading to frequency conversion of the input pump into spectral anti-Stokes and Stokes sidebands with large (i.e., > 100 THz) frequency detuning in the visible and infrared domains, respectively. The next nonlinear process is the stimulated Raman scattering (SRS) cascade up to 1.3 – 1.4 μm, followed by soliton propagation and associated self-frequency shift [24-25, 30-32, 33]. Note that the generated multimode solitons feature very high (up to the MW range) peak powers, which permits to drastically improve the quality of the nonlinear imaging of biological samples via two and three-photon absorption, when using infrared light. We would like to underline that the output beam pattern was found to remain highly robust against environmental perturbations, e.g., fibre bending and squeezing.

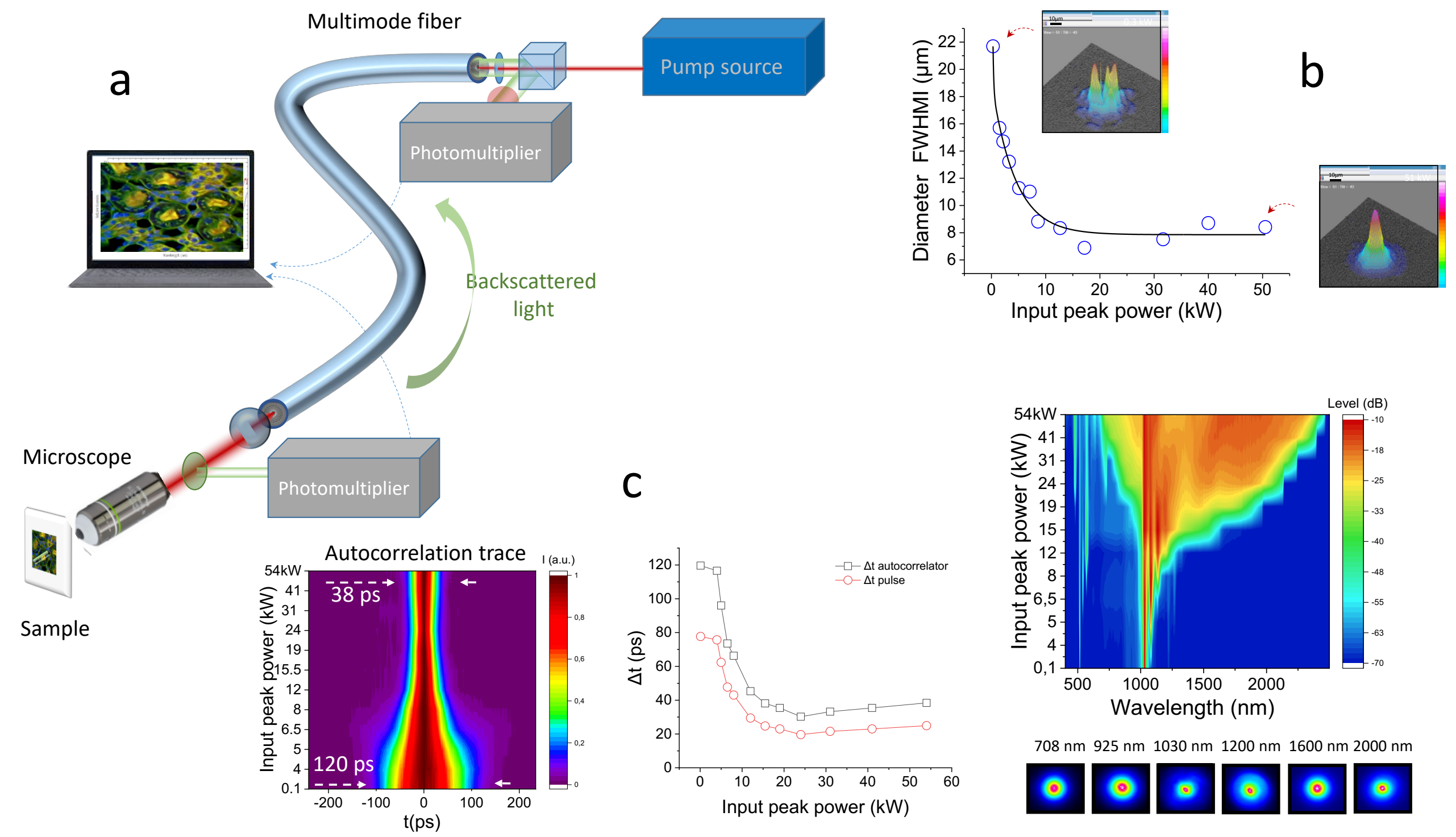

**Figure 1: Experimental setup and characteristics of the output self-cleaned beam.** a, Schematic representation of the experimental setup as detailed in Methods section. b, Output beam diameter evolution vs. input peak power (Insets: 3D representation of the output beam pattern for low and high peak powers, respectively). c, Output autocorrelation (left) and pulse duration (right) versus the input peak power. d, supercontinuum generation vs. input peak power (insets: near field pattern of the output beam for different wavelengths). Optical 50/125 µm GRIN fibre, 3 m for BSC, 18 m for supercontinuum generation.

Next, we used the self-cleaned beam at 1064 nm to perform nonlinear fluorescence imaging of kidney of mouse and bovine endothelial cells (see Supplementary). The total average power sent on the biological sample was limited to a few 5 mW with 200 kHz repetition rate. In these conditions, we obtained images of tubule, actin and nucleus, labelled with Alexa 488, Alexa 568 and Dapi, respectively. The absorption mechanism is two-photon fluorescence for Alexa fluorochromes, and three-photon fluorescence for Dapi (see figure 3a).

In order to demonstrate the advantage of our set-up, we compared the stability of nonlinear fluorescence imaging in two different situations. Specifically, when using a self-cleaned pump beam, or a speckled pump beam with the same average power. We realized intensity correlations between two images taken with two-photon fluorescence. The biological sample was the kidney of mouse (Invitrogen FluoCells n°3 F24630), illuminated at 1064 nm, and analyzed at 640 nm, in order to observe actin labelled with Alexa Fluor 568. All of the images were taken with a continuously squeezed coiled optical fibre, in order to vary the random mode coupling process inside the fibre. In the first step, and by using the spatial self-cleaning process (see Fig.2a), the intensity correlation is mainly distributed along the diagonal direction, which demonstrates a strong replication of images taken from the same sample, but with a delayed time. The estimated correlation coefficient is close to 98.5 %, by using imageJ correlator. To the contrary, two different images recorded with a speckled beam at the fibre output undergo significant distortions, as it can be clearly seen on the correlation diagram of Fig.2b. In this case, the image correlation is also mainly distributed along the diagonal axis, but with a significant spreading, as testified the below 90% correlation coefficient (it can reach even lower values, depending on the strength of the externally imposed fibre squeezing).

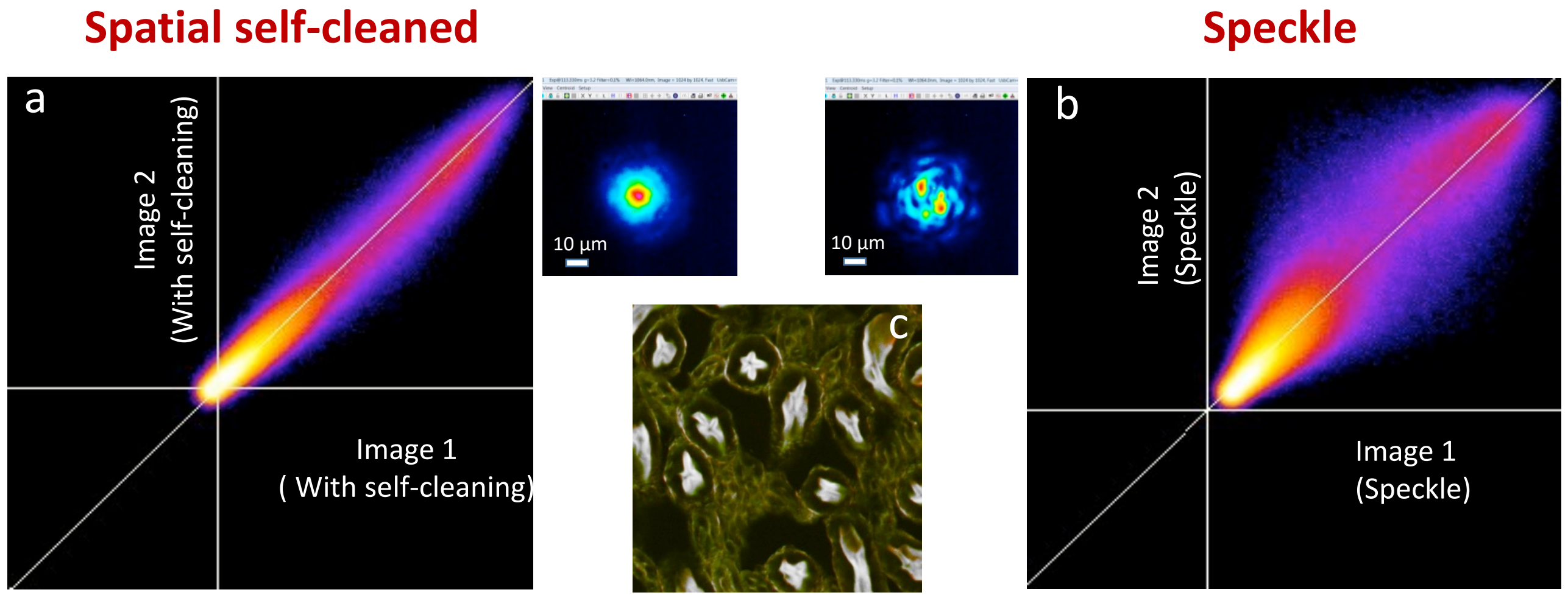

**Figure 2: Illustration of self-cleaning robustness and image stability when continuous squeezing is applied to the multimode fibre.** a, Intensity correlation between two different images, recorded with a self-cleaned output beam at 1064 nm. b, Intensity correlation between two different images, recorded with a speckled output beam. Insets: 2D images of the output beams with (Left) and without (Right) spatial self-cleaning. Example of images used for intensity correlation. Pump wavelength 1064 nm, dwell time: 5µs, repetition rate: 200 kHz, average power on the sample: 5 mW.

Besides image stability, we also investigated the impact of temporal pulse reshaping accompanying BSC. Using beams that propagate in the GRIN fibre in a linear regime, the signature of the Dapi fluorochrome remained low, and no image of nucleus could be obtained. The same experiment was performed with a self-cleaned beam, by keeping the average power on the sample unchanged. This time, a clear signature of nucleus was obtained via three-photon fluorescence imaging, thanks to the increased pulse peak power, hence demonstrating the significant beneficial impact of BSC (see Supplementary).

By increasing the peak power coupled in the GRIN fibre, we could generate a supercontinuum ranging from 700 nm to 2.4 µm. By using an optical filter (see Methods), we sliced from the supercontinuum a wavelength range in the near-infrared domain between 700 and 950 nm, and performed nonlinear fluorescence imaging. Images were obtained with two-photon absorption fluorescence (see figure 3b). As previously remarked, all of the images remained strongly stable with respect to fibre perturbations. This is because the supercontinuum was carried by a robust bell-shaped spatial beam across its entire wavelength range.

We performed equivalent experiments in the infrared domain, between 1300 nm and 1500 nm, and at the specific wavelength of 1200 nm, to illuminate Dapi fluorochrome by three-photon absorption. Multimode solitons exist for wavelengths longer than range of 1300 nm - 1350 nm, which contains the zero-dispersion wavelength (ZDW) for the low-order modes that carry the main portion of the output power. The average power sent on the sample remained the same as in previous experiments. The obtained images (see figure 3c) lead us to conclude that operating in the multimode soliton regime significantly improves the efficiency of the nonlinear imaging process.

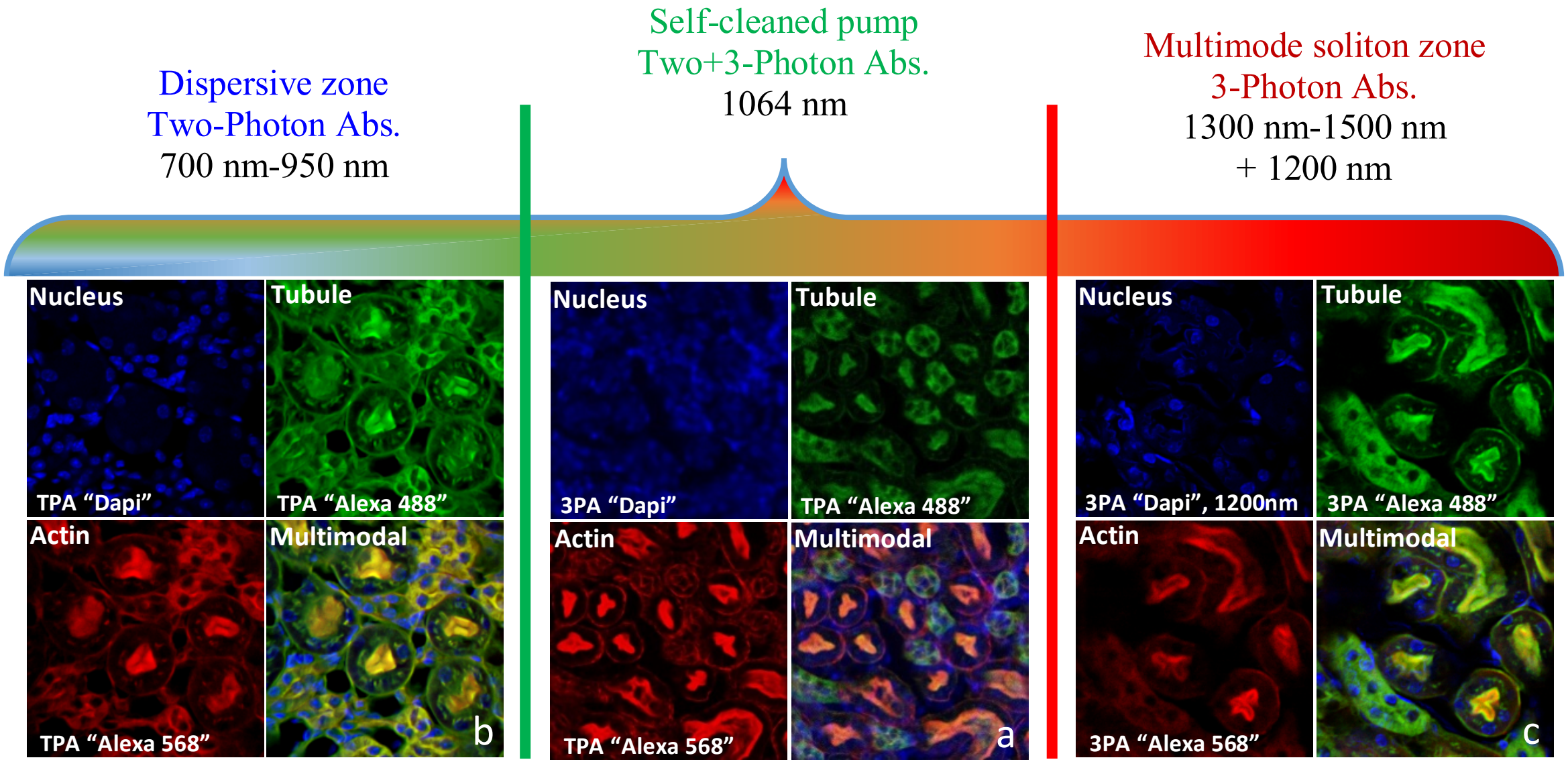

**Figure 3: Nonlinear fluorescence images obtained in a microscopy configuration of mouse kidney labelled with Alexa 488, Alexa 568 and Dapi, revealing tubule, actin and nucleus, respectively.** a, two-photon fluorescence imaging of tubule and actin, three-photon imaging of nucleus by using a self-cleaned pump beam at 1064 nm. b, two-photon fluorescence imaging of tubule, actin and nucleus by using visible/IR light between 700 nm and 950 nm. c, three-photon fluorescence imaging of tubule, actin and nucleus by using IR light between 1300 nm and 1500 nm and 1200 nm for Dapi; Dwell time: 5µs/px, averaged traces for 1 image: 20, image size: 1024×1024 pixels.

To quantify the advantages of our new imaging system, we determined its spatial 3D and 2D resolutions. The results are shown in figure 4. The self-cleaned beam at 1064 nm permits transverse and longitudinal resolutions of 0.66 µm and 3.1 µm respectively (see figure 4a and 4b): these results represent a significant improvement with respect to the linear multimodal regime, where the multiple focal points, caused by the initial speckled beam, greatly weaken the spatial resolution beyond 1.5 µm. The 3D PSF is shown figure 4c. When using near infrared (750-950 nm) and infrared (1300-1500 nm) light, the spatial quality of the output beam is slightly improved, because of SRS beam clean-up and nonlinear conversion via dispersive wave generation. By using self-cleaned beams, we can reach the maximum resolution allowed by our system, i.e., 0.37 µm and 0.54 µm in the near-infrared and infrared zone, respectively (see figure 4d, and 4f). Importantly, all studied configurations allowed to clearly resolve the sub-nuclear structure, such as nucleolus, and mitochondrial network as microtubules (see figure 4e).

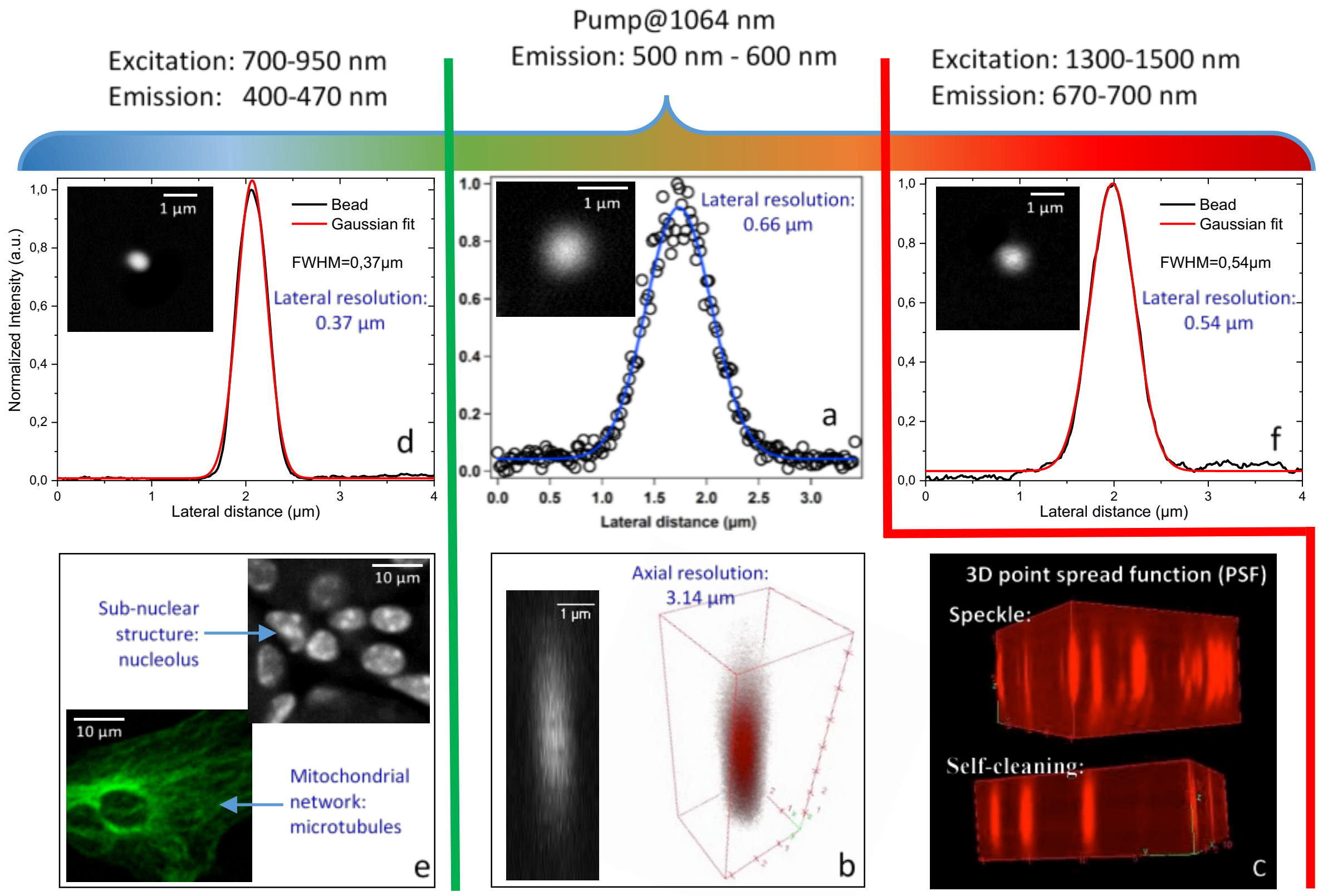

**Figure 4: Measure of the spatial resolution of our fluorescence imaging microscopy system based on two-photon absorption, for three different wavelength domains.** a, transverse resolution at the input pump wavelength; b, axial resolution at the pump wavelength; c, 3D PSF for a speckled beam (top) or a self-cleaned beam (bottom) at 1064 nm. d, lateral resolution in the range 700-950 nm; e, example of high-resolution images showing sub-nuclear structure (nucleolus) and microtubules; f, lateral resolution in the infrared domain (1300-1500 nm).

Finally, in the last experiment, we studied the possibility to obtain nonlinear fluorescence images in the endoscopic configuration. First, we used the pump beam wavelength to record two-photon fluorescence of actin and tubule, labelled with Alexa 568 and Alexa 488, respectively (see figure 5a and 5b). We measured the evolution, versus input peak power, of fluorescence on cellulose microfibril, in order to confirm the multiphoton nature of the nonlinear absorption process (see fig. 5f). By selecting the infrared part of the supercontinuum beyond 1.55 µm (see fig. 5e), we obtained three-photon fluorescence imaging of cellulose microfibril (see fig. 5d, to be compared with two-photon imaging in fig. 5c). When operating in the anomalous dispersion regime, multimode soliton propagation drastically increases the pulse peak power, thus facilitating high-order multiphoton nonlinear imaging. However, in contrast with the microscopy configuration, in the endoscopic configuration the image contrast of actin and tubule was reduced. This could be due to nonlinear conversion into the visible region via supercontinuum generation, which cannot be filtered out; no filter can be used after the fibre, because of the backward fluorescence signal, which is detected after its backward propagation through the fibre. The use of shorter fibres could mitigate the generation of geometric parametric instabilities, and improve the signal-to-noise ratio.

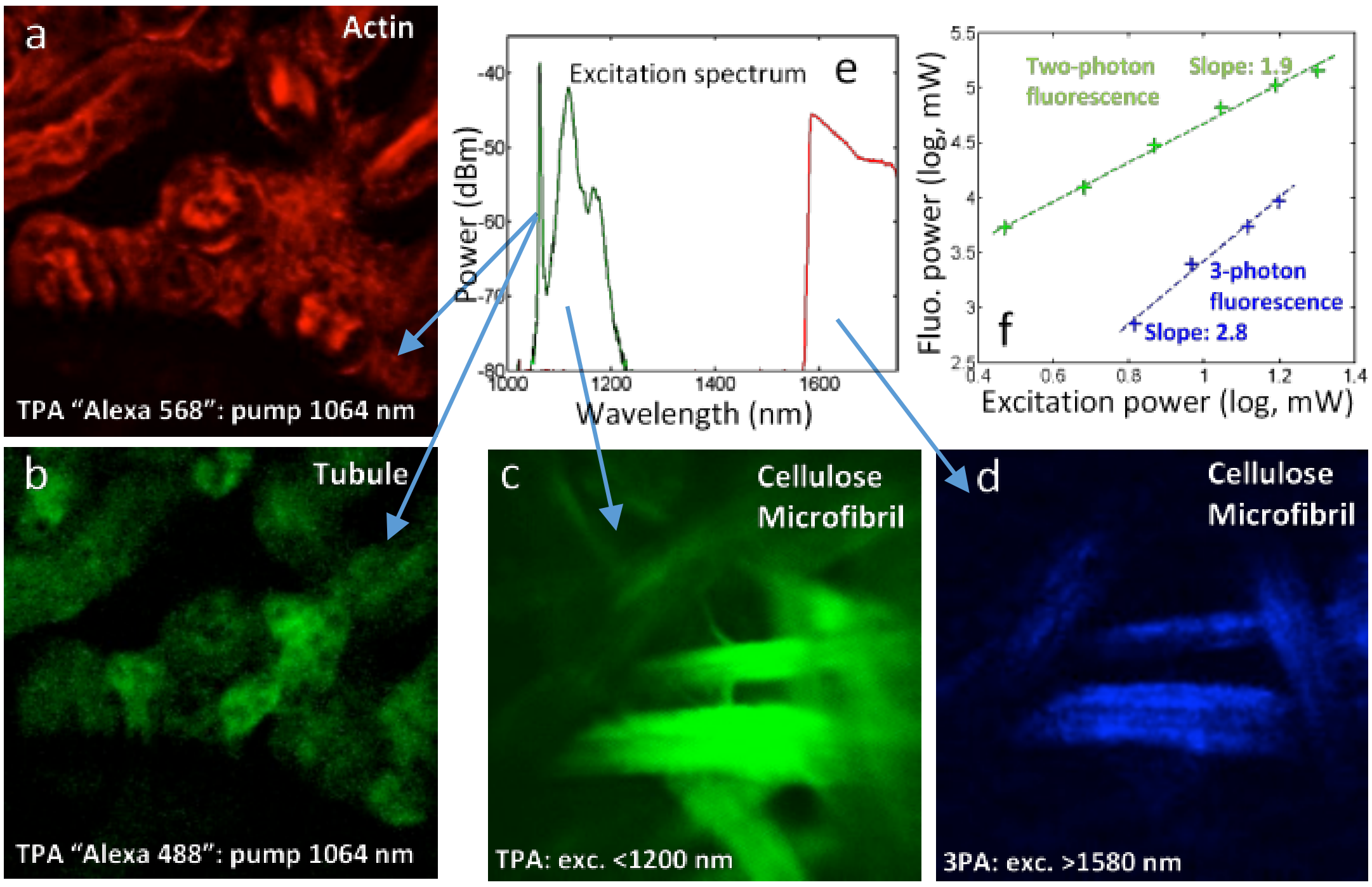

**Figure 5: Nonlinear fluorescence images, obtained in endoscopy configuration, of mouse kidney labelled with Alexa 488 and Alexa 568, revealing tubule and actin.** a, b, two-photon fluorescence images of actin and tubule by using the self-cleaned pump beam at 1064 nm. c, d, two and three-photon fluorescence images of cellulose microfibril; e, output spectrum used for experiments; f, fluorescence evolution versus input peak power.

In summary, we experimentally demonstrated that ultrafast spatiotemporal BSC based on nonlinear optical pulse propagation in a GRIN fibre can be exploited for high-resolution, and environmentally robust MMF-based nonlinear fluorescence imaging of biological samples. Temporal pulse narrowing associated with BSC was found to increase the output pulse peak power, thus strongly enhancing the efficiency of multiphoton imaging. This especially occurs with a pump wavelength beyond 1.3 µm, because of the generation of high-energy multimode solitons. Spectral broadening, associated with spatiotemporal beam cleaning, can be optimized to cover the entire spectral range spanning from the visible to the infrared domain, by generating a supercontinuum carried by a cleaned spatial profile, further allowing for nonlinear imaging in a wideband configuration. The other key feature of our system is the high stability of the imaging measurement, which opens the way for the development of high resolution, environmentally-robust microscopic and endoscopic fluorescence imaging systems based on MMFs. The capability to provide both high peak and average power is another advantage of employing MMFs, which can pave the way for multispectral LIDAR applications. Our results provide the building blocks for harnessing the complexity of nonlinear spatiotemporal multimode dynamics, and developing novel bio-imaging techniques with greatly improved performances.

## Methods

**Microscopy experiments.** We used a mode-locked and amplified pump laser, delivering Fourier transformed pulses (80 ps) at 1064 nm with adjustable repetition rate up to 2 MHz and a peak power up to 1 MW. The infrared linearly-polarized Gaussian beam was coupled in a 50/125 GRIN MMF by using 45 mm converging lens. Its core diameter was 52.1 µm, with a numerical aperture of 0.205. The focused input beam had a 35 µm diameter at full width at half the maximum intensity (FWHMI). Thus, 99% of the guided input power was coupled into about 80 modes. The angle injection was properly adjusted to be 0°, to favour self-cleaning process on the fundamental mode. We also used a polarizer cube placed in between two half-wave plates, in order to control the beam power while adjusting the input beam polarization orientation. With a 3 m long fibre span, we obtained BSC without any significant spectral broadening [4]. In a second step, by increasing both the fibre length (18 m) and the input peak power (up to 50 kW), spectral broadening of the input pump pulse was obtained, until a supercontinuum was generated, covering the full wavelength range between 600 nm and 2.5 µm. At the fibre output, a converging lens of 4.5 mm focal length and a half-wave plate, collimated the output beam and adjusted its polarization orientation. An adjustable neutral density filter was placed at the fibre output, in order to control the power sent into the microscope after propagation in the GRIN MMF. Three different filters, coupled with the appropriate control of the input peak power, were used to select the desired wavelength bands: a bandpass filter FL-1064-10 (Thorlabs), a longpass filter FELH1300 (Thorlabs) and a shortpass filter FESH950 (Thorlabs), were used to select the spectral content of the pump beam at 1064 nm, 700-950 nm and 1300-1500 nm ranges, respectively. A longpass filter FELH1200 (Thorlabs) was used to select wavelengths close to 1200 nm, in order to excite the nucleus labelled with Dapi. Next, the filtered beam was sent in an upright multiphoton microscope (Bergamo-Thorlabs) with galvanometric scanners and two photomultipliers. In order to focus the incident beam on the sample, we used two microscope objectives (Olympus, XLUMPLFLN20XW and LUMFLN60XW). Three additional single-band bandpass filters at 460/80 nm, 525/50 nm and 607/70 nm from Semrock (brightline) were inserted in front of the photomultipliers, in order to select the corresponding fluorescence emission. An afocal system composed of two converging lenses was introduced on the initial laser beam path, in order to obtain a spatial magnification of the incident beam, thus allowing for covering of the entire input window of the objective, and reaching the maximum spatial resolution. Images with 1024 × 1024 pixels were recorded with a dwell time of 5 µs/px with twenty accumulations.

**Endoscopy experiments.** The laser source used for the endoscopy experiments was the same as that used in the microscopy configuration. BSC was also obtained in the same conditions. An additional longpass filter at 700 nm was placed at 45° before the input fibre end, in order to reflect the backward fluorescence generated in the samples towards two additional photomultipliers. A shortpass filter at 750 nm (FES750, Thorlabs) and a longpass filter at 450 nm (FEL450, Thorlabs) were used to select the corresponding fluorescence band in the front of the photomultipliers.

**Biological samples for microscopy experiments.** We used multilabeled cell preparations from mouse kidney section provided by Invitrogen (FluoCells n°3 F24630), and labelled with Alexa 488, Alexa 568 and Dapi. A second sample was bovine pulmonary artery endothelial cells (FluoCells n°2 F14781) labelled with Dapi, Bodipy FL and Texas Red (the corresponding results are displayed in the supplementary material). Two additional samples of cellulose microfibril were used for the endoscopy experiments.

**Acknowledgements**
M. F., M. S., A.T., and V.C. acknowledge the financial support provided by: the French ANR through the "TRAFIC project: ANR-18-CE080016-01"; the CILAS Company (ArianeGroup) through the shared X-LAS laboratory; the "Région Nouvelle Aquitaine" through the projects F2MH, SIP2 and Nematum; the National Research Agency under the Investments for the future program with the reference *ANR-10-LABX-0074-01 Sigma-LIM*. K. K. acknowledges the Foundation of Polish Science (TEAM-NET project No. POIR.04.04.00-00-16ED/18-00); M.P., V.C., M.F., M.Z., F.M., A.N., D.M., L.L. and S. W. acknowledge the European Research Council (ERC) under the European Union's Horizon 2020 research and innovation programme (No. 875596, No. 740355). S.W. acknowledges the Italian Ministry of University and Research (R18SPB8227) and the Russian Ministry of Science and Education (14.Y26.31.0017); G. M. acknowledges the Conseil Régional de Bourgogne Franche-Comté, the iXcore research foundation and the National Research Agency (ANR-15-IDEX-0003, ANR-17-EURE-0002).

**Author contributions**
N. O. M., T. M., C-H. A., M. F. and V. C. carried out the experiments. All authors analysed the obtained results, and participated in the discussions and in the writing of the manuscript.

**Additional information**

**Competing financial interests**
The authors declare no competing financial interests.